\documentstyle[pra,aps,epsf,eqsecnum]{revtex}

\begin{document}
\def\beq{\begin{equation}}
\def\eeq{\end{equation}}
\def\bea{\begin{eqnarray}}
\def\eea{\end{eqnarray}}
\def\oupb{UPB\ }
\def\pb{PB\ }
\def\eps{\epsilon}
\newcommand{\ket}[1]{| #1 \rangle}
\newcommand{\bra}[1]{\langle #1 |}
\newcommand{\braket}[2]{\langle #1 | #2 \rangle}
\newcommand{\proj}[1]{| #1\rangle\!\langle #1 |}
\newcommand{\ba}{\begin{array}}
\newcommand{\ea}{\end{array}}
\newtheorem{theo}{Theorem}
\newtheorem{defi}{Definition}
\newtheorem{lem}{Lemma}
\newtheorem{cor}{Corollary}
\newtheorem{exam}{Example}
\newtheorem{prop}{Proposition}
\newcommand{\abs}[1]{ \mid #1\mid}
\newcommand{\av}[1]{\langle\,#1\,\rangle}
\newcommand{\PT}[1]{#1^{T_2}}
\newcommand{\ngeq}{\not\ge}
\author{David P. DiVincenzo$^*$, 
Peter W. Shor$^\ddag$,
John A. Smolin$^*$, Barbara M. Terhal$^*$, and Ashish V. Thapliyal$^\dag$}

\title{Evidence for Bound Entangled States with Negative Partial Transpose}

\address{\vspace*{1.2ex} \hspace*{0.5ex}{$^*$ IBM T.J. Watson Research
Center, Yorktown Heights, NY 10598 USA, 
$^\ddag$AT\&T Research, Florham Park, NJ 07932 USA,
$^\dag$Dept. of Physics, University of California, Santa Barbara,
California 93106 USA.\\ }}

\date{\today}

\maketitle
\begin{abstract}
We exhibit a two-parameter family of bipartite mixed states
$\rho_{bc}$, in a $d\otimes d$ Hilbert space, which are negative under
partial transposition (NPT), but for which we conjecture that no
maximally entangled pure states in $2\otimes 2$ can be distilled by
local quantum operations and classical communication (LQ+CC).
Evidence for this undistillability is provided by the result that, for
certain states in this family, we cannot extract entanglement from
any arbitrarily large number of copies of $\rho_{bc}$ using a
projection on $2\otimes 2$.  These states are canonical NPT states in
the sense that any bipartite mixed state in any dimension with NPT
can be reduced by LQ+CC operations to an NPT state of the $\rho_{bc}$
form.  We show that the main question about the distillability of
mixed states can be formulated as an open mathematical question about
the properties of composed positive linear maps.

\end{abstract}

%

\section{Introduction}

Maximally entangled quantum states, when their two halves are shared
between two parties, are a uniquely valuable resource for various
information-processing tasks.  Used in conjunction with a quantum
communications channel, they can increase the classical data carrying
capacity of that channel, in some cases by an arbitrarily large
factor\cite{C_E}.  Possession of maximally entangled states can ensure
perfect privacy of communication between the two parties by the use of
quantum cryptography\cite{Ek1}.  These states can facilitate the rapid
performance of certain forms of distributed computations\cite{Clev}.
Of course, maximally entangled states are the key resource in quantum
teleportation\cite{tele}.  On the other hand, the surreptitious
establishment of entanglement between two parties can thwart the
establishment of trust between parties via bit commitment\cite{MLC}.

How can two parties come into the possession of a shared maximally
entangled state?  If the storage and transportation of quantum
particles were perfect, then the state could have been synthesized in
some laboratory long in the past and given to Alice and Bob (our
personified parties) for storage until needed.  In practice no such
perfect infrastructure exists.  Since the most interesting scenarios
for the use of quantum entanglement are in cases where Alice and Bob
are remote from one another, we will consider the long-distance
transportation of quantum states needed to establish the shared
entanglement to be difficult and imperfect, while the local processing
of quantum information (unitary transformations, measurement) we will
assume, for the sake of analysis, to be essentially perfect.

Under these assumptions, when we wish to assess whether a given
physical setup is or is not useful for entanglement assisted
information processing, our analysis focuses on the mixed quantum
state, $\rho$, in the hands of Alice and Bob after the difficult
transportation step.  We enquire whether $\rho^{\otimes n}$ can be
transformed, by {\em LQ+CC operations}, to a supply of maximally
entangled states.  Here the $\otimes n$ notation indicates that $n$
copies of the state $\rho$ are available, and we will be concerned
with asymptotic results as $n$ is taken to infinity.  LQ+CC operations
(sometimes called LOCC in the literature) are obtained by any
arbitrary sequence of {\em local} quantum {operations} (appending
ancillae, performing unitary operations, discarding ancillae)
supplemented by {\em classical communication} between Alice and Bob.

An interesting fact about this possibility for the {\em distillation}
of entanglement is that it is neither rare nor ubiquitous; a finite
fraction of the set of all possible bipartite mixed states $\rho$ can
be successfully distilled\cite{filterhor}, and a finite fraction
cannot\cite{Brau}.  Much work has been focussed on whether $\rho$
falls into the distillable or into the undistillable class, and this
paper is primarily a contribution to this classification task.  Before
describing our new contributions, we will give a brief review of
previous results on classifying states according to their
distillability.

\begin{figure}
\epsfxsize=15cm
\epsfbox{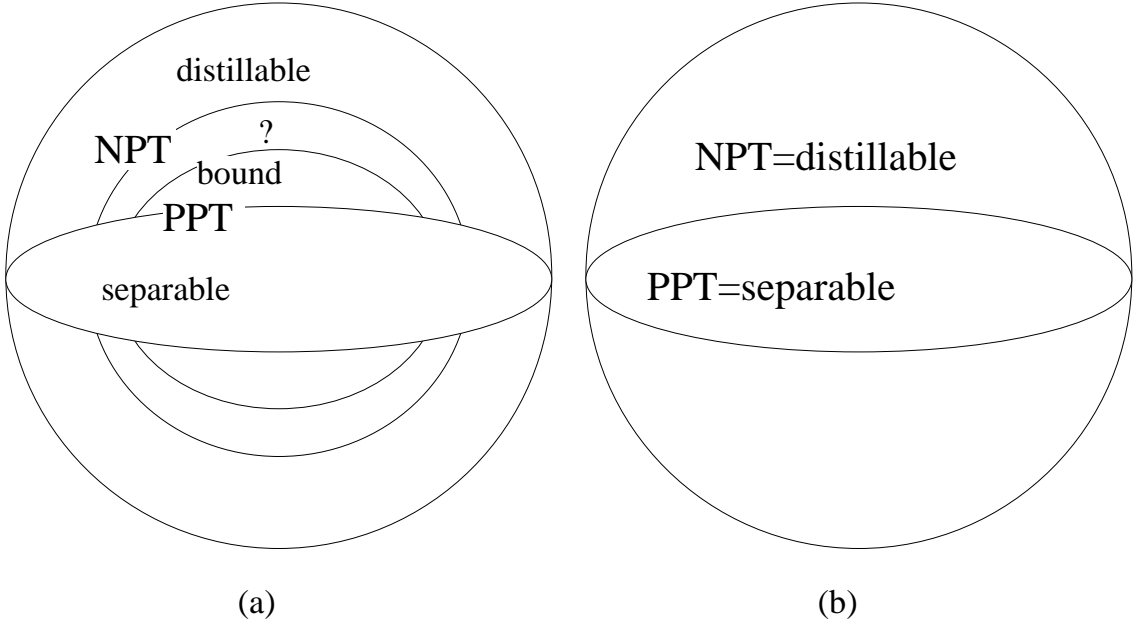}
\caption{Layout of the set of all mixed states.  (a) General case for
arbitrary Hilbert space dimension $m\otimes n$.  The `?' region, that
of bound or undistillable NPT states, is the subject of this paper.
This region is known to contain no states for $2\otimes n$. (b) Simplified
situation for dimension of $2\otimes 2$ and $2\otimes 3$ for which it
is know that all PPT states are separable, and all NPT states are
distillable.}
\label{PPTetc}
\end{figure}

Multipartite density matrices $\rho$ are considered {\em unentangled}
if there exists a decomposition of $\rho$ into an ensemble of pure
product states; for the bipartite case this means that we can write
\begin{equation}
\rho=\sum_i p_i |\alpha_i\rangle\langle\alpha_i|\otimes
|\beta_i\rangle\langle\beta_i|.\label{defsep}
\end{equation}
These are also referred to as separable states.  It is clear that
separable states are never distillable.  However, the converse
proposition, that entangled states are always distillable, is false in
general, although true for density matrices in $2\otimes 2$ and
$2\otimes 3$ Hilbert spaces\cite{foot1}.  This became clear shortly
after the introduction by Peres\cite{PerLet} of a computationally
simple criterion for separability, the {\em partial transposition}
test.  The partial-transpose operation, denoted as $I\otimes T$ when
the transpose is applied to Bob's Hilbert space, is specified by the
action
\begin{equation}
\langle ij|(I\otimes T)(\rho)|kl\rangle=\langle il|\,\rho\,|kj\rangle.
\end{equation}
While application of $T$ to Alice's Hilbert space will lead to
identical results, we will always apply it to Bob's space in this
paper.  Here $\ket{i}$ and $\ket{k}$ indicate an orthonormal basis on
Alice's Hilbert space, and $\ket{j}$ and $\ket{l}$ the same for Bob.
It is easy to show that separable states are positive under partial
transpose, that is, that the matrix
\begin{equation}
(I\otimes T)(\rho)=\rho^{PT}
\label{intronot}
\end{equation}
is a positive semi-definite operator, denoted by $\rho^{PT}\ge 0$.
(Eq. (\ref{intronot}) introduces the $PT$ notation that we will use
throughout this paper.)  This positivity property is abbreviated as
PPT; states for which $\rho^{PT}\ngeq 0$ are called NPT states.  It
was soon recognized\cite{phoro} that the set of PPT density operators
$\rho$ is larger than the set of unentangled states (except in
$2\otimes 2$ and $2\otimes 3$); see Fig. \ref{PPTetc}.  It was also
discovered that all PPT states, even those which are inseparable, are
not distillable.  The existence of such states, in which entanglement
is present (since entanglement is required to synthesize the states)
but cannot be re{\"e}xtracted in pure form, was a surprising
observation, indicating the possibility of a fundamentally new form of
irreversibility in physics.  States having this property are said to
possess {\em bound entanglement}.

The introduction of the PPT/NPT classification suggested a new
conjecture about distillability, namely that all states with NPT would
possess distillable entanglement, and it is the purpose of the present
paper to explore this conjecture.  While no rigorous results have been
obtained concerning this conjecture, we will introduce a two-parameter
family of NPT states for which we obtain evidence that the conjecture
is false.  That is, we consider it likely that the family of states we
introduce below has only bound entanglement, despite being NPT.

We have been able to recast the question about the distillability
property of the $\rho_{bc}$ states, or of any NPT states, as a
question about the two-positivity properties of certain positive
linear maps\cite{nec_horo}.  These maps arise because there is a
one-to-one correspondence between mixed states on $d\otimes d$ and
completely positive linear maps ${\cal S}$ on $d$ dimensions which,
when applied to half a maximally entangled state $\ket{\Psi^+}$,
produces the mixed state $\rho$.  Then, the NPT property is related to
the map $T\circ{\cal S}$ and its compositions $(T\circ{\cal
S})^{\otimes n}$.  The open question about distillability can be posed
compactly as a question concerning the mathematical properties of
these maps.  This approach also permits us to consider the question of
whether distillability is an additive property, that is, whether the
amount of distillable entanglement of $\rho_1\otimes\rho_2$ is just
the sum of the two separately.  Horodecki {\em et al.}\cite{active}
have given some evidence for a kind of undistillability involving
single copies of PPT bound entangled states.  In the positive-map
language, the most general questions about non-additivity can be
compactly framed.  This shows that further developments of the theory
of positive maps will be very desirable in settling some of the
fundamental questions about the entanglement properties of quantum
states.

This paper develops in the following way: Section \ref{canon}
introduces the canonical states $\rho_{bc}$ and shows the LQ+CC
mapping that produces them.  Section \ref{tools} considers the
distillability of any state $\rho$ by application of the basic
criterion of whether it remains entangled when projected into
$2\otimes 2$.  Section \ref{single} considers a single copy of the
$\rho_{bc}$ state, establishing the $\rho_{bc}$ for which there exist
such projections into $2\otimes 2$.  Sec. \ref{multi} takes up the
much harder case of multiple copies, with Sec. \ref{nulls} proving the
result that for some $\rho_{bc}^{\otimes n}$ states, no entanglement
remains upon projection into $2\otimes 2$ even for arbitrarily large
$n$.  Section \ref{twopos} recasts the question about distillability
in terms of two-positivity of linear maps isomorphic to the mixed
states.

\section{A canonical set of NPT density matrices}
\label{canon}

The desired, but too-ambitious, program would be to assess the
distillability of all NPT states.  We will attempt this assessment
only for a specific subset of the NPT states parameterized by two real
numbers.  This subset will, however, have a specific relation to the
set of NPT states, in that there is a LQ+CC operation that will map
the general NPT state onto one in our two-parameter family.  This
LQ+CC operation preserves the NPT property.  Thus, if we could exhibit
a protocol for the distillation of our two-parameter family, this
would suffice to show that all NPT states were distillable.  On the
contrary, our canonical two-parameter family has properties which make
distillation quite hard for certain ranges of the parameters,
suggesting that in fact some portion of the full set of NPT states is
{\em not} distillable.

Our canonical states, with real parameters $b$ and $c$, are written as
\begin{equation}
\rho_{bc}=a \sum_{i=0}^{d-1} \ket{ii} \bra{ii} +
b\sum_{i,j=0,i < j}^{d-1}\ket{\psi_{ij}^-} \bra{\psi_{ij}^-}+
c\sum_{i,j=0,i < j}^{d-1} \ket{\psi_{ij}^+} \bra{\psi_{ij}^+}.
\label{abcform}
\end{equation}
Here
\begin{equation}
\ket{\psi_{ij}^{\pm}}=\frac{1}{\sqrt{2}}(\ket{ij} \pm \ket{ji}).
\end{equation}
The states live in a $d\otimes d$ Hilbert space.  The parameter $a$ in
Eq. (\ref{abcform}) is not independent, because of the unit trace
condition it is related to $b$ and $c$ by
\begin{equation}
d a+(b+c)d(d-1)/2=1.
\end{equation}

The range of interest for the parameters $b$ and $c$ is shown in
Fig. \ref{NPTreg}.  As we will show in the next section, the state is
NPT in two triangular regions of parameter space; one of these regions
${\rm NPT}_2$, which will not be of much interest to us (all these
states are distillable), lies above the straight line $KJ$, and is
defined by the inequality $c>2/d^2+b(d-2)/2$.  The region ${\rm
NPT}_1$, about which we will have much more to say, lies in the region
$BFK$ and is defined by $b>1/(d(d-1))$.  Region $ABKJ$ contains PPT
states; in Sec. \ref{multi} we prove that all these states are also
separable.

To show that $\rho_{bc}$ represents a canonical set, we will exhibit a
procedure involving only LQ+CC operations that will convert any NPT
density matrix $\rho$, that is, one satisfying the condition
\begin{equation}
\langle \psi |\, ({\bf 1} \otimes T) (\rho)\, | \psi \rangle  < 0,
\label{negfirst}
\end{equation}
for some state $\ket{\psi}$, to one of the $\rho_{bc}$ form having
NPT.  We will take the Hilbert space dimension to be $n\otimes m$,
that is, we will not restrict Alice's and Bob's dimensions to be the
same.

Here is the sequence of LQ+CC operations that will reduce the general
NPT state $\rho$ to $\rho_{bc}$:

(i) {\em rotation to the Schmidt basis:} We write the $\ket{\psi}$ of
Eq. (\ref{negfirst}) as
\begin{equation}
\ket{\psi}=\sum_{i=0}^{d-1} \sqrt{\lambda_i} \ket{\alpha_i} \otimes
\ket{\beta_i}.
\end{equation}
Here $d\leq \min(n,m)$.  Let $U_A \ket{\alpha_i}=\ket{i}$ and $U_B
\ket{\beta_i}=\ket{i}$, or
\begin{equation}
\ket{\psi}=U_A^{\dagger} \otimes U_B^{\dagger} \ket{\phi},
\end{equation}
where
\begin{equation}
\ket{\phi}=\sum_{i=0}^{d-1} \sqrt{\lambda_i} \ket{i} \otimes \ket{i}.
\end{equation}
We define $\rho^{(i)}=
U_A \otimes U_B\,\rho\,U_A^{\dagger} \otimes U_B^{\dagger}$.
Equation (\ref{negfirst}) can be rewritten as
\begin{equation}
\langle \phi |\,({\bf 1} \otimes T^U) (\rho^{(i)})\,| \phi \rangle < 0.
\label{negin}
\end{equation}
where $T^U$ is transposition in a rotated basis determined by $U_B$.
The negativity of the expression Eq. (\ref{negin}) does not depend on the
basis in which $T$ is performed, therefore we will replace $T^U$ by $T$
again in the remainder.

(ii) {\em local filtering (see \cite{filterhor}):}
We define the state $\ket{\Phi^+}$ as
\begin{equation}
\ket{\Phi^+}=\frac{1}{\sqrt{d}}\sum_{i=0}^{d-1} \ket{i} \otimes \ket{i}.
\end{equation}
The filter operation $W$ on Alice's Hilbert space is defined by the
equation
\begin{equation}
W^\dagger \otimes {\bf 1} \ket{\Phi^+}=\ket{\phi}.
\end{equation}
We apply this local filter to the state $\rho^{(i)}$ to obtain $\rho^{(ii)}$:
\begin{equation}
\rho^{(ii)}=\frac{(W \otimes {\bf 1})\, \rho^{(i)}\,
(W^{\dagger} \otimes {\bf 1})}
{{\rm Tr}\, (W^{\dagger}W \otimes {\bf 1})\, \rho^{(i)}}.
\end{equation}
Eq. (\ref{negfirst}) implies that
\begin{equation}
\bra{\Phi^+}\, ({\bf 1} \otimes T) (\rho^{(ii)})\, \ket{\Phi^+}=
{\rm Tr}\, \ket{\Phi^+} \bra{\Phi^+}\, ({\bf 1} \otimes T) (\rho^{(ii)}) < 0.
\label{NPTc1}
\end{equation}
We now use that ${\rm Tr}\, (A^\dagger\, T(B))={\rm Tr} (T^{\dagger}
(A^\dagger)\, B)$ and $T^{\dagger}=T$ to rewrite this NPT condition in
a form which will be convenient below:
\begin{equation}
{\rm Tr}\,H \rho^{(ii)} < 0,
\label{neg2}
\end{equation}
with
\begin{equation}
H=({\bf 1} \otimes T) (\ket{\Phi^+} \bra{\Phi^+}).\label{NPTc2}
\end{equation}
This Hermitian operator $H$ can be written in its eigenbasis:
\begin{equation}
H=\frac{1}{d}\sum_{i=0}^{d-1} \ket{ii} \bra{ii}
-\frac{1}{d} \sum_{i,j=0,i < j}^{d-1} \ket{\psi_{ij}^-} \bra{\psi_{ij}^-}
+\frac{1}{d} \sum_{i,j=0,i < j}^{d-1}
\ket{\psi_{ij}^+} \bra{\psi_{ij}^+},\label{NPTc3}
\end{equation}
where
\begin{equation}
\ket{\psi_{ij}^{\pm}}=\frac{1}{\sqrt{2}}(\ket{ij} \pm \ket{ji}).
\end{equation}

(iii) {\em project into $d\otimes d$:} Since $\ket{\Phi^+}$, and $H$,
have support only a $d\otimes d$ dimensional subspace of the Hilbert
space, Alice and Bob can project locally onto this subspace and leave
the NPT condition Eq. (\ref{NPTc1}), or Eq. (\ref{NPTc2}), unchanged.
We call the resulting NPT density matrix in $d\otimes d$ $\rho^{(iii)}$.

(iv) {\em diagonal twirl:} Alice and Bob perform a equal mixture of
identical unitary operations, which are diagonal in the Schmidt basis
given by the vectors $\ket{i}$, giving state $\rho^{(iv)}$.  This
unitary operation is
\begin{equation}
(U_{A,B}(\{\theta\}))_{i,j}=\delta_{ij}e^{i\theta_i}.
\end{equation}
The phases $\theta_i$ are chosen randomly over a uniform distribution
from 0 to $2\pi$, independently for each $i$.  This leaves the
operators
\begin{equation}
\ket{ij} \bra{ji}, \ket{ij} \bra{ij}, \ket{ii}\bra{ii}
\end{equation}
invariant. This operation therefore leaves the eigenvectors of $H$ and
thus $H$ itself invariant. Thus it follows that
\begin{equation}
{\rm Tr}\,H \rho^{(iv)}={\rm Tr}\,\int d\{\theta\}
U^\dagger(\{\theta\})\otimes U^\dagger(\{\theta\})
HU(\{\theta\})\otimes U(\{\theta\})\rho^{(iii)}=
{\rm Tr}\,H\rho^{(iii)}<0.
\end{equation}
The `twirled' density matrix $\rho^{(iv)}$ has the form:
\begin{equation}
\rho^{(iv)}=\sum_{i=0}^{d-1} \alpha_i \ket{ii} \bra{ii}+
\sum_{i,j=0,i \neq j}^{d-1}\beta_{ij}^1 \ket{ij} \bra{ij}+
\sum_{i,j=0,i \neq j}^{d-1}\beta_{ij}^2  \ket{ij} \bra{ji}.
\end{equation}
Note that the coefficients in these sums are all in general
distinct, with $\beta_{ij}^1$ not necessarily equal to $\beta_{ji}^1$ and
similarly for $\beta_{ij}^2$.

(v) {\em symmetrize by permutation:} Alice and Bob carry out identical,
randomly chosen unitary transformations which are drawn uniformly from
all possible permutation operations over the elements of the Schmidt
basis $\ket{i}$.  This ensures that in the new density matrix
$\rho^{(v)}$ the $\alpha_i$ coefficients, for all $i$, become equal to
a single number $a$, all the $\beta_{ij}^1$ become equal (we call this
constant $c+b\over 2$), and all the $\beta_{ij}^2$ become equal (we call this
constant $c-b\over 2$).  So we obtain
\begin{equation}
\rho^{(v)}=a\sum_{i=0}^{d-1} \ket{ii} \bra{ii}+
{c+b\over 2}\sum_{i,j=0,i \neq j}^{d-1} \ket{ij} \bra{ij}+
{c-b\over 2}\sum_{i,j=0,i \neq j}^{d-1} \ket{ij} \bra{ji}.
\end{equation}
But comparing with Eq. (\ref{abcform}), we note that we have arrived
at the desired canonical form,
\begin{equation}
\rho^{(v)}=\rho_{bc}.
\end{equation}
As the Hermitian matrix $H$ of Eqs. (\ref{NPTc2}) and (\ref{NPTc3}) is
again invariant under this symmetrization, we note that the NPT property
is again preserved:
\begin{equation}
{\rm Tr}H \rho^{(v)}={\rm Tr}H\rho_{bc}<0.
\end{equation}

We may summarize the foregoing line of argument as a Theorem:

\begin{theo}
Let $\rho$ be a bipartite density matrix on $n \otimes m$ with the
property that $\rho^{PT}\ngeq 0$.  The density matrix $\rho$ can be
converted by local operations and classical communication to a density
matrix $\rho_{bc}$ on $d\otimes d$ with $d\leq\min(n,m)$ characterized
by two real parameters $b$ and $c$ such that $\rho^{PT}_{bc}\ngeq 0$.
This density matrix $\rho_{bc}$ is
\begin{equation}
\rho_{bc}=a \sum_{i=0}^{d-1} \ket{ii} \bra{ii} +
b\sum_{i,j,i < j}^{d-1}\ket{\psi_{ij}^-} \bra{\psi_{ij}^-}+
c\sum_{i,j,i < j}^{d-1} \ket{\psi_{ij}^+} \bra{\psi_{ij}^+},
\label{abcform2}
\end{equation}
with
\begin{equation}
d a+(b+c)d(d-1)/2=1.\label{constr}
\end{equation}

\end{theo}

It is easy to see from the form of $H$ that these transformations
carry all NPT states $\rho$ into a $\rho_{bc}$ sitting in the ${\rm
NPT}_1$ region of Fig. \ref{NPTreg}.  This is why the ${\rm NPT}_2$
region will not be of concern to us.

We note that it is possible to follow the five-step reduction above
with another LQ+CC operation, resulting in a canonical NPT density
operator characterized by just a single real parameter:

(vi) {\em full twirl:} Alice and Bob perform a equal mixture of
identical unitary operations drawn uniformly (with the Haar measure)
from the entire group $U(d)$.  It is straightforward to show that
the resulting density matrix $\rho^{(vi)}$ has the same form as above
(Eq. (\ref{abcform2})):
\begin{equation}
\rho^{(vi)}=a' \sum_{i=0}^{d-1} \ket{ii} \bra{ii} +
b'\sum_{i,j,i < j}^{d-1}\ket{\psi_{ij}^-} \bra{\psi_{ij}^-}+
c'\sum_{i,j,i < j}^{d-1} \ket{\psi_{ij}^+} \bra{\psi_{ij}^+},
\label{abcform3}
\end{equation}
with
\begin{eqnarray}
&&b'=b,\\
&&c'={2\over{d(d+1)}}-{{d-1}\over{d+1}}b,
\end{eqnarray}
and $a'$ given by the same constraint as in Eq. (\ref{constr}).  Thus,
$\rho^{(vi)}$ depends only on the single parameter $b$; it is the same
Werner density matrix studied recently by Horodecki {\em et
al.}\cite{filterhor}:
\begin{equation}
\rho_W={1\over{d^3-d}}[(d-\phi){\bf 1}+(d\phi-1)dH],
\label{Werner}
\end{equation}
note that $H$ of Eq. (\ref{NPTc2}) is proportional to the ``swap''
operator
\begin{equation}
dH\ket{i}\otimes\ket{j}=\ket{j}\otimes\ket{i}.
\end{equation}
This ``full twirl'' carries all the states in the $BFK$ region of
Fig. \ref{NPTreg} onto the line $FH$, without changing the value of
$b$.

Of course, if it were possible to prove that all the NPT states of the
one-parameter form $\rho_W$ were distillable, then all NPT states
would be distillable through the reductions we have developed above.
In fact we conjecture, as Horodecki {\em et al.} have previously
(Sec. VIII, Ref. \cite{filterhor}), that some of these NPT states are
undistillable.  Under these circumstances, it is desirable to provide
evidence for undistillability for the widest class of states possible,
and we will concentrate in this paper on providing such evidence for
the two-parameter family of canonical states $\rho_{bc}$, more
particularly, for those lying near the line segment $BK$ in
Fig. \ref{NPTreg}.  All of the results we develop will, of course,
also apply to the restricted one-parameter family $\rho_W$ as well.

\section{Tools for the study of distillability}
\label{tools}

In this section we will explore all the known tools at our disposal
for analyzing the distillability of states.  For some of the
$\rho_{bc}$ states we believe that no distillation protocol exists;
evidence for this is provided by the last result of this section, that
for some $\rho_{bc}$ states, any successful distillation protocol, if
it exists, must act on some very large number $n$ of copies of the
state; we show that $n$ must diverge along an entire boundary $BK$ in
Fig. \ref{NPTreg}.

Much of the discussion of distillation strategies will need the
notion of the {\em Schmidt rank} of a pure state in an ensemble
decomposition of density matrix $\rho$.  We first define this term:

\begin{defi}
A bipartite pure state $\ket{\psi}$ has Schmidt rank k if the state can
be written in the Schmidt polar form as
\begin{equation}
\ket{\psi}=\sum_{i=1}^{k} \sqrt{\lambda_i} \ket{a_i} \otimes \ket{b_i},
\end{equation}
with $\langle a_i | a_j \rangle=\delta_{ij}$ and $\langle b_i|b_j
\rangle=\delta_{ij}$.
\end{defi}

The distillation of the $\rho_{bc}$ states (or more particularly, of
the $\rho_W$ subset of these states) has already been considered
in\cite{filterhor}.  There, a distillation protocol was developed
based on the positive linear map $\Lambda_c \colon \rho \rightarrow
{\rm Tr}\rho {\bf 1} -\rho$.  In Sec. \ref{twopos} we will discuss
other aspects of the relation between the theory of positive maps and
the distillability of mixed states.  For all states $\rho$ for which
$({\bf 1}\otimes\Lambda_c)(\rho)\ngeq 0$, it was shown how to distill
them by converting these states to a different canonical
density-matrix form introduced by Werner.

However, all the states $\rho_{bc}$ remain positive under the action
of $\Lambda_c$, so long as the dimension $d>2$, because $({\bf
1}\otimes\Lambda_c)(\rho_{bc})\propto\rho_{b'c'}$, where
$b'=(c+b)/2-b/(d-1)$ and $c'=(c+b)/2-c/(d-1)$.  (Positivity under the
action of $\Lambda_c$ was already known for $\rho_W$\cite{filterhor}.)
Thus, the simple distillation procedure studied in\cite{filterhor}
will not work for these states.  Thus, to study the distillability of
these states, we need to consider the more general necessary and
sufficient condition developed by Horodecki {\em et al.}:

\begin{lem} (Horodecki et al.\cite{horodeckibound2}) A density matrix
$\rho\in m_A\otimes m_B$ is distillable if and only if there exists
a finite $n$ and projections $P_A\colon {\cal
H}_{m_A}^{\otimes n} \rightarrow {\cal H}_2$ and $P_B\colon {\cal
H}_{m_B}^{\otimes n} \rightarrow {\cal H}_2$ such that
$\sigma=(P_A \otimes P_B)\, \rho^{\otimes n}\, (P_A^{\dagger} \otimes P_B^{\dagger})$
is entangled.
\label{horlem}
\end{lem}
In $2\otimes 2$, a density matrix $\sigma$ is entangled if and only
if it is NPT.

Lemma \ref{horlem} requires the examinations of projection of the
density matrix (or $n$ copies of the density matrix).  The following
Lemma gives a convenient recasting of these properties of projections
in terms of properties of the original density matrix itself:

\begin{lem}
\label{x2n}
Let $\rho$ be a density matrix on $m_A\otimes m_B$.  Let $P_A\colon
{\cal H}_{m_A}\rightarrow{\cal H}_2$ be a projection and also
$P_B\colon {\cal H}_{m_B}\rightarrow{\cal H}_2$.  There exist $P_A$
and $P_B$ such that $P_A\otimes P_B\,\rho\, P^\dagger_A \otimes
P^\dagger_B$ is entangled if and only if
\begin{equation}
\rho_{2 \otimes m_B}=P_A \otimes {\bf 1}_B
\rho P^\dagger_A \otimes {\bf 1}_B
\label{oneside}
\end{equation}
has the property that
\begin{equation}
\rho^{PT}_{2\otimes m_B}\ngeq 0.
\label{notPSD}
\end{equation}
Eq. (\ref{notPSD}) is equivalent to the condition that there exists a
state $\ket{\phi}$ that has Schmidt rank two and
\begin{equation}
\bra{\phi}\,({\bf 1} \otimes T) (\rho)\,\ket{\phi} < 0.
\label{negexp}
\end{equation}
\end{lem}

{\em Proof:} If the density matrix $\rho_{2 \otimes m_B}$ is not
positive semidefinite under partial transposition, then there exists a
Schmidt rank two vector $\ket{\psi}$, written in its Schmidt basis as
\begin{equation}
\ket{\psi}=\sqrt{\lambda_1}\ket{a_0,b_0}+\sqrt{\lambda_2}\ket{a_1,b_1},
\label{sd}
\end{equation}
such that
\begin{equation}
\bra{\psi}\, \rho_{2 \otimes m_B}^{PT}\, \ket{\psi} < 0.
\label{2nneg}
\end{equation}
(The state $\ket{\psi}$ cannot be a product vector since, if it were,
$\bra{\psi}\, \rho_{2 \otimes m_B}^{PT}\, \ket{\psi}={\rm Tr}\, 
\ket{\psi}\bra{\psi}\,\rho_{2 \otimes m_B}^{PT}={\rm Tr}\,
(\ket{\psi}\bra{\psi})^{PT}\,\rho_{2 \otimes m_B} \geq 0$.)

We note that the projector $P_A$ in Eq. (\ref{oneside}) consistent with
Eq. (\ref{sd}) has the form
$P_A=\ket{a_0}\bra{a_0}+\ket{a_1}\bra{a_1}$.  Note also
that the state $\ket{\psi}$ is invariant under the projector
$P_B=P_B^\dagger=\ket{b_0}\bra{b_0}+\ket{b_1}\bra{b_1}$,
\begin{equation}
({\bf 1}_A \otimes P^\dagger_B)\ket{\psi}=\ket{\psi}.\label{inv}
\end{equation}
Plugging Eqs (\ref{inv}) and (\ref{oneside}) into Eq. (\ref{2nneg}):
\begin{equation}
\bra{\psi}({\bf 1}_A\otimes P_B) [(P_A \otimes {\bf 1}_B)\, \rho\,
(P^\dagger_A \otimes {\bf 1}_B)]^{PT}({\bf 1}\otimes P^\dagger_B)\ket{\psi}=
\bra{\psi} [(P_A \otimes P^*_B)\, \rho\, (P^\dagger_A \otimes P^T_B)]^{PT}
\ket{\psi}< 0.\label{long1}
\end{equation}
Therefore the state $(P_A\otimes P^*_B)\, \rho\,(P^\dagger_A \otimes
P^T_B)$ on $2\otimes 2$ is entangled.

Conversely, if the density matrix $\rho_{2 \otimes m_B}$ is positive
semidefinite under partial transposition for all $P_A$, meaning that
$\rho_{2 \otimes m_B}$ is either separable or has bound entanglement,
then there does not exist a $P_B$ such that $(P_A \otimes P_B)\,\rho\,
(P^\dagger_A\otimes P^\dagger_B)$ is entangled, because then it could be
distilled.

Finally, by rewriting Eq. (\ref{long1}) as
\begin{equation}
\bra{\psi} (P_A \otimes P_B)\, \rho^{PT}\, (P^\dagger_A \otimes P^\dagger_B)
\ket{\psi}< 0,\label{long}
\end{equation}
we note that $\ket{\phi}=(P^\dagger_A \otimes P^\dagger_B)\ket{\psi}$
is the state needed for Eq. (\ref{negexp}).  $\Box$

Note that an easy consequence of Lemma \ref{x2n} is that all NPT
states in $2\otimes n$ for any $n$ are distillable.

\subsection{Single copy}
\label{single}

The real difficulty in applying Lemma \ref{horlem} is that it requires
the examination of an arbitrary number of copies $n$ of the state
to be distilled.  We will therefore first develop a set of strong
results for the special case of $n=1$, then we will move on to
obtain some results for the much more difficult case of arbitrary
$n$.

We begin with some terminology:

\begin{defi}
{\rm : We say that density matrix $\rho$ is {\em pseudo one-copy
undistillable} if, for all Schmidt rank two states $\ket{\phi}$,
$\bra{\phi}\,\rho^{PT}\,\ket{\phi}\ge 0$.  Then, by Lemma \ref{x2n},
there exists no $2\otimes2$ projection of $\rho$ that is inseparable.
We say $\rho$ is {\em pseudo $n$-copy undistillable} if and only if
$\rho^{\otimes n}$ is pseudo one-copy undistillable.}
\end{defi}

\begin{figure}
\epsfxsize=15cm
\epsfbox{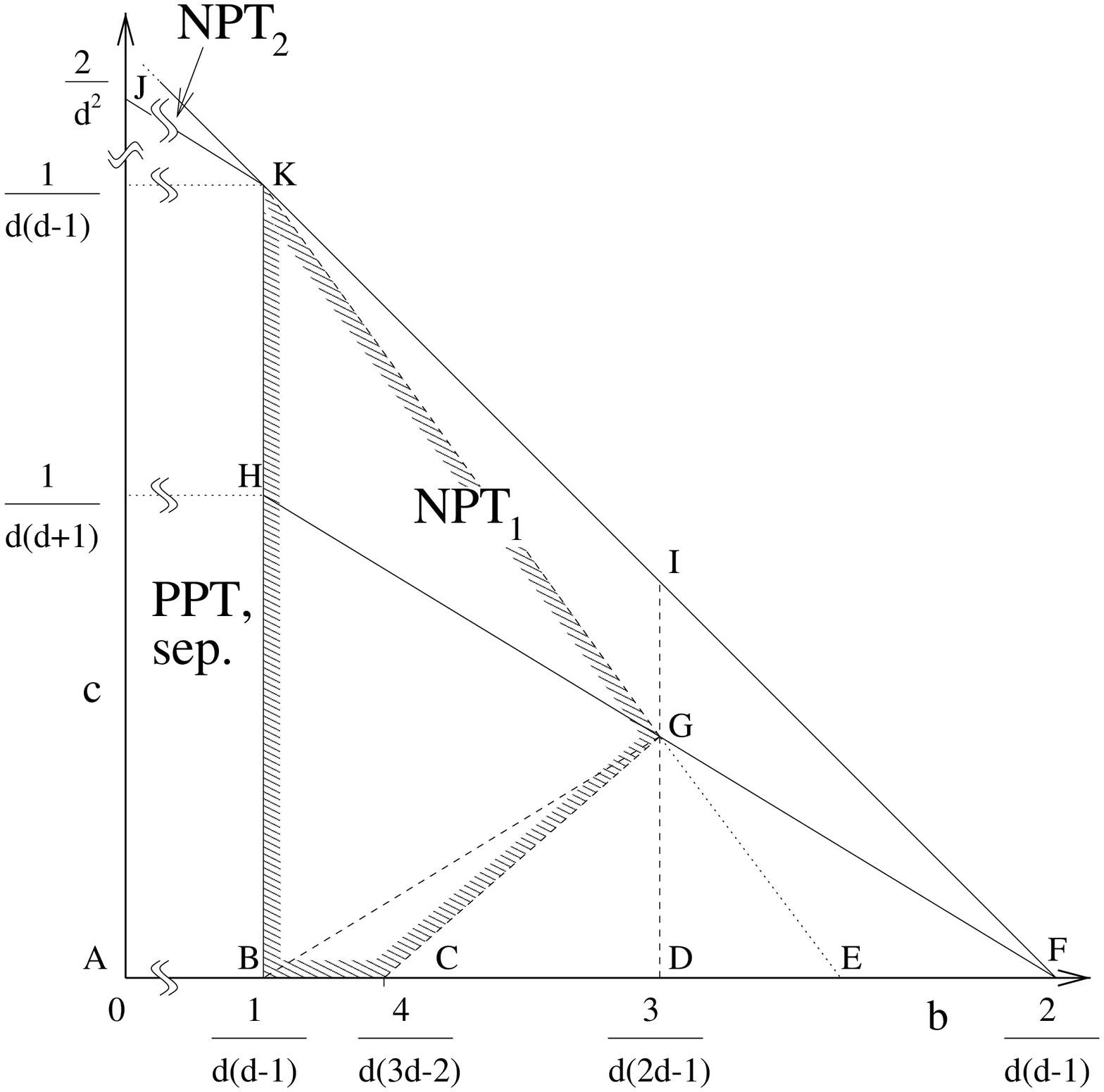}
\caption{The relevant region of the $bc$ parameter space for the
states $\rho_{bc}$.  All NPT states can be brought by LQ+CC action
into the region ${\rm NPT}_1$, triangle $BFK$.  For general dimension,
region $CFKG$ is distillable by projection on one copy and region
$BCGK$ is pseudo one-copy undistillable.  In $3\otimes 3$ we have
strong evidence that region $BCGK$ is pseudo two-copy undistillable.
We conjecture that the entire region $BCGK$ is undistillable by any
means.  All states in the PPT region $ABKJ$ are separable; that is,
there are no bound PPT states among the $\rho_{bc}$ set.}
\label{NPTreg}
\end{figure}

We will establish which states $\rho_{bc}$ are pseudo one-copy
undistillable and which are distillable.  The partial transpose of
$\rho_{bc}$ reads
\begin{equation}
\rho^{PT}_{bc}=a \sum_{i=0}^{d-1} \ket{ii}\bra{ii} +
\frac{c-b}{2} \sum_{i,j=0; i\ne j}^{d-1}
\ket{ii}\bra{jj} + \frac{c+b}{2} \sum_{i,j=0;i\ne j}^{d-1} \ket{ij}\bra{ij}.
\label{abcpt}
\end{equation}
The eigendecomposition of $\rho^{PT}_{bc}$ is
\begin{equation}
\rho^{PT}_{bc} = \lambda_0 \proj{\Phi_0} +
\lambda_1 \sum_{i=1}^{d-1} \proj{\Phi_i} +
\lambda_2 \sum_{i,j=0; i\ne j}^{d-1} \proj{ij},
\label{eigvecs1}
\end{equation}
with
\begin{equation}
\ket{\Phi_k} = \frac{1}{\sqrt{d}}\sum_{j=0}^{d-1}e^{i 2 \pi j k/d}\ket{jj},
\label{eigvecs2}
\end{equation}
which we refer to as the ``e-dit eigenstates'' in analogy with
``ebit'', because they are the maximally entangled states in $d\otimes
d$ having a ``dit''($\log_2 d $ bits) of entanglement.
Correspondingly, we refer to the $\ket{ij}$ states with $i\ne j$ as
the ``product eigenstates''. The eigenvalues $\lambda_i$ are given by
\begin{eqnarray}
&&\lambda_0=(d-1)\left({1\over{d(d-1)}}-b\right)\,\,\,\,
(<0\mbox{ in NPT}_1),\\
&&\lambda_1={1\over d}-{d\over 2}c-{{d-2}\over 2}b\,\,\,\,
(>0\mbox{ in NPT}_1),\\
&&\lambda_2={1\over 2}(c+b)\ge 0.
\end{eqnarray}
The negative eigenvalue $\lambda_0$ is independent of $c$, showing why
the PPT-NPT boundary is a vertical line ($BK$ in Fig \ref{NPTreg}).
Notice that the eigenvectors of $\rho^{PT}_{bc}$ are independent of
parameters $b$ and $c$.

We now specialize to the state for which the positive eigenvalues are
all equal, $\lambda_1=\lambda_2$, and therefore
\begin{equation}
c={2\over{d(d+1)}}-{{d-1}\over{d+1}}b.
\end{equation}
These are precisely the Werner states $\rho_W$ of Eq. (\ref{Werner})
above, the states along the line $FH$ in Fig. \ref{NPTreg}.  We take
advantage of the fact that Lemma \ref{x2n} does not require normalized
states to write the partial transpose of these states in the following
simple unnormalized form:
\begin{equation}
\label{eq:sigmaPT}
\sigma^{PT}(\lambda)=\lambda I - (\lambda+1)\proj{\Phi_0} \enspace,
\label{ldef}
\end{equation}
with $\lambda=\lambda_1/(-\lambda_0)$.  We will show that for
$\lambda\ge 2/(d-2)$,
$\min_{\ket{\psi^2}}\bra{\psi^2}\,\sigma^{PT}\,\ket{\psi^2}\ge 0$ and
that for $\lambda< 2/(d-2)$,
$\min_{\ket{\psi^2}}\bra{\psi^2}\,\sigma^{PT}\,\ket{\psi^2}<0$, with
the minimum taken over all Schmidt rank two vectors.  Thus
$\lambda=2/(d-2)$, corresponding to $b=3/(d(2d-1))$ and
$c=1/(d(2d-1))$ (the point $G$ in Fig. \ref{NPTreg}) is the transition
point separating distillable Werner states (line segment $FG$) from
those which are pseudo one-copy undistillable (line segment $GH$).  To
establish this we first need to prove the following Lemma:
\begin{lem}
\label{lem:overlapschmidt}
In $d\otimes d$, the overlap of a Schmidt rank two state with a
maximally entangled state is at most $\sqrt{2/d}$.  In other words, if
$\ket{v}$ has Schmidt rank two and $\ket{\Psi}$ is a maximally
entangled state, then
\begin{equation}
|\braket{\Psi}{v}|\le\sqrt{2/d}\, .
\end{equation}
\end{lem}
{\em Proof:} In its Schmidt basis,
$\ket{\Psi}=(\sum_{i=0}^{d-1}\ket{ii})/\sqrt{d}$. Since $\ket{v}$ is
Schmidt rank two, it may be written in its Schmidt decomposition as
$\ket{v}=\sqrt{\mu_1}\ket{e_1}\ket{e_2}+\sqrt{\mu_2}\ket{e_3}\ket{e_4}$,
with $\mu_1+\mu_2=1$. The overlap then is,
\begin{eqnarray}
\braket{\Psi}{v} & = &
\frac{\sqrt{\mu_1}}{\sqrt{d}} \sum_{i=0}^{d-1} \braket{i}{e_1} \braket{i}{e_2}
+\frac{\sqrt{\mu_2}}{\sqrt{d}} \sum_{i=0}^{d-1} \braket{i}{e_3} \braket{i}{e_4}
\nonumber \\
& = &\frac{\sqrt{\mu_1}}{\sqrt{d}} \sum_{i=0}^{d-1}
\braket{i}{e_1} \braket{e_2^*}{i}
+\frac{\sqrt{\mu_2}}{\sqrt{d}} \sum_{i=0}^{d-1}
\braket{i}{e_3} \braket{e_4^*}{i}
\nonumber \\
&=& \frac{\sqrt{\mu_1}}{\sqrt{d}} \braket{e_2^*}{e_1} +
\frac{\sqrt{\mu_2}}{\sqrt{d}} \braket{e_4^*}{e_3},
\end{eqnarray}
where $\ket{e_i^*}$ is the vector obtained by complex conjugation of
the components of $\ket{e_i}$ in the Schmidt basis of the state
$\ket{\Psi}$.  Thus, we have
\begin{equation}
|\braket{\Psi}{v}|  \le  \frac{\sqrt{\mu_1}+\sqrt{\mu_2}}{\sqrt{d}} \enspace.
\end{equation}
Maximizing with constraint $\mu_1+\mu_2=1$ gives the desired result.
$\Box$

Now we are ready for the main result:
\begin{theo}
\label{1shot}
Given $\sigma(\lambda)$ whose partial transpose is given in
Eq. (\ref{eq:sigmaPT}), we have,
\begin{itemize}
\item if $\lambda\ge 2/(d-2)$ then $\sigma$ is not pseudo one-copy distillable.
\item if $\lambda< 2/(d-2)$ then $\sigma$ is pseudo one-copy distillable.
\end{itemize}
\end{theo}
{\em Proof:} We start with the first part. Let $\ket{v}$ be any Schmidt
rank two vector. Then,
\begin{eqnarray}
\bra{v}\sigma^{PT}\ket{v}&=&
\lambda-(\lambda+1)\abs{\braket{v}{\Phi_0}}^2\nonumber\\
& \ge & \lambda - 2(\lambda+1)/d \nonumber \\
& \ge & {d-2\over d}\left(\lambda-{2\over d-2}\right)\, ,
\end{eqnarray}
where we have used Lemma \ref{lem:overlapschmidt}. This is greater
than or equal to zero for $\lambda\ge 2/(d-2)$, showing the first part
of the result.  For the second part, consider
$\ket{v}=(\ket{00}+\ket{11})/\sqrt{2}$. We have
$\bra{v}\sigma^{PT}\ket{v}=((d-2)/2)(\lambda-2/(d-2))$, which is less than
zero for $\lambda< 2/(d-2)$, proving the second part of the result.
$\Box$

From this it is a simple matter to completely characterize the
one-copy undistillability of the $\rho_{bc}$ states:

\begin{prop}
\label{abc1shotno}
The states $\rho_{bc}$ are pseudo one-copy undistillable in the region
of parameter space BCGK in Fig. \ref{NPTreg}.\cite{momo}
\end{prop}

{\em Proof:} Since any state in the region is a convex linear
combination of the states $B$, $C$, $G$, and $K$, it suffices to show
that the partial transpose of each of these four states has a positive
expectation value with respect to any Schmidt rank two vector (Lemma
\ref{x2n}).  This is obviously true for the PPT states $B$ and $K$,
and it is true for state $G$ by Theorem \ref{1shot}.  To show it for
$C$, which has parameters $b=4/(d(3d-2))$, $c=0$, we note that the
partial transpose of the state $C$ can be written
\begin{equation}
\rho^{PT}\left(b={4\over d(3d-2)},c=0\right)=
{2d-1\over 3d-2}\rho^{PT}_G+{2\over
d(3d-2)}\sum_{i,j=0,i<j}^{d-1}\Pi_{ij}.\label{deco}
\end{equation}
Here $\rho^{PT}_G$ is the partial transpose of the normalized state at
point $G$, and $\Pi_{ij}$ is the normalized projector
$\Pi_{ij}={1\over 2}(\ket{ii}-\ket{jj})(\bra{ii}-\bra{jj})$.  The
expectation value of the first term on the right-hand side of
Eq. (\ref{deco}) is positive by Theorem \ref{1shot}, and that of the
second term is positive because it is a projector.$\Box$
\medskip

All other states are distillable:

\begin{prop}
\label{abc1shot}
The states $\rho_{bc}$ are distillable in the region of parameter
space CFKG in Fig. \ref{NPTreg}.\cite{momo}
\end{prop}
{\em Proof:} In the region $EFK$ the partial transpose has a negative
expectation value with respect to the Schmidt rank two state
$\ket{00}+\ket{11}$, and in the region $CEG$ with respect to the
state
\begin{equation}
\left(\sum_{j=0}^{d-1}\ket{j}\right)\otimes\left(\sum_{k=0}^{d-1}\ket{k}
\right)+
\left(\sum_{j=0}^{d-1}e^{2\pi ij/d}\ket{j}\right)\otimes\left(\sum_{k=0}^{d-1}
e^{2\pi ik/d}\ket{k}\right).\ \ \Box
\end{equation}

\subsection{Multiple copies}
\label{multi}

It has proved to be much harder to obtain definitive results concerning
the pseudo $n$-copy undistillability of the $\rho_{bc}$ states.  But we
have accumulated various pieces of evidence, which we will present here,
all indicating the likelihood that many of the ${\rm NPT}_1$ states
are undistillable.

Our attention will focus here on a particular subset of the
$\rho_{bc}$ states labeled by $c$ and a small parameter $\eps$, which
sit just to the right of the line segment $BK$ in Fig. \ref{NPTreg}:
\begin{equation}
\rho(c,\eps)=\left({1\over 2d}-{d-1\over 2}(c+\eps)\right)
\sum_{i=0}^{d-1} \ket{ii} \bra{ii}+
\left({1\over d(d-1)}+\eps\right)
\sum_{i < j} \ket{\psi_{ij}^-} \bra{\psi_{ij}^-}+
c \sum_{i < j} \ket{\psi_{ij}^+} \bra{\psi_{ij}^+}.
\label{epsstates}
\end{equation}
The eigenvectors of the partial transpose of this state
$\rho(c,\eps)^{PT}$ are given in Eqs. (\ref{eigvecs1}-\ref{eigvecs2}),
since these are common to all $\rho_{bc}$ states.  The eigenvalues are
$\lambda_0=-(d-1)\eps$, $\lambda_1={1\over{2(d-1)}}-{d-2\over
2}\eps-{d\over 2}c$ and $\lambda_2={1\over d}\left(
{1\over{2(d-1)}}+{d\over 2}\eps+{d\over 2}c\right)$.  The only
properties of these eigenvalues that we will use is that for small,
positive $\eps$ and $0\le c<1/(d(d-1))$, $\lambda_0$ is negative and
goes to zero as $\epsilon\rightarrow 0$, and $\lambda_1$ and
$\lambda_2$ are strictly positive.

Although we will not need any more properties of the density matrices
$\rho(c,\eps=0)$, we can at this point note the interesting fact that
they are all separable; in fact, {\em all} the PPT states of the
form $\rho_{bc}$ (the region $ABKJ$ in Fig. \ref{NPTreg}) are
separable (Eq. (\ref{defsep})).  This is established by showing that
the density matrices at the extremal points $A$, $B$, $K$, and $J$ are
separable; all other states in this region are convex combinations of
these.  The state at $A$ is proportional to $\sum_i\ket{ii}$, and the
one at $K$ is proportional to $\sum_{i\neq j}\ket{ij}$, so these are
both obviously separable.

We can also create the state $\rho(c=0,\eps=0)$ at point $B$ using
separable states.  It is easiest to construct this ensemble for the
partial transpose of this state (see Eq. (\ref{abcpt})), which is done
by equally mixing the states
\begin{equation}
(-\ket{i}+e^{2 \pi i k/3} \ket{j}) \otimes (\ket{i}+e^{-2\pi i k/3} \ket{j}),
\end{equation}
for all pairs $i\neq j$, and $k=0,1,2$.  By mixing these states with
equal probabilities, all terms of the form $\ket{ii}\bra{ij}$,
$\ket{ij}\bra{ii}$ and $\ket{ij}\bra{ji}$ for $j \neq i$ cancel out;
each of these will come with a factor $\sum_{k=0}^2e^{\pm 4\pi i
k/3}=0$ or $\sum_{k=0}^2e^{\pm 2\pi i k/3}=0$.  A term such as
$\ket{00}\bra{00}$ will occur $d-1$ times as much as a term
$\ket{00}\bra{11}$, which is indeed the correct ratio for
$\rho(c=0,\eps=0)^{PT}$.  The state $\rho(c=0,\eps=0)$ itself at point
$B$ is obtained from mixing the states
\begin{equation}
(-\ket{i}+e^{2 \pi i k/3} \ket{j}) \otimes (\ket{i}+e^{
2\pi i k/3} \ket{j})\label{dec1}
\end{equation}
with equal probabilities.

The partial transpose of the state at point $J$ has a simple form
($\lambda_1=0$ in Eq. (\ref{eigvecs1})); it is straightforward
to show that $\rho^{PT}$ at $J$ is realized by an equal mixture of
the separable states
\begin{equation}
(\sum_{j=0}^{d-1}e^{2\pi ik_j/3}\ket{j})\otimes
(\sum_{j=0}^{d-1}e^{-2\pi ik_j/3}\ket{j}),
\end{equation}
where each integer $k_0$, $k_1$, ... $k_{d-1}$ runs independently
over 0, 1, and 2.  This is clearly not a separable decomposition
with the minimal possible number of states.

A few notes about the decomposition for point $B$: for $d=3$ the state
$\rho^{PT}$ at point $B$ has rank eight.  This implies that the
optimal decomposition of $\rho^{PT}$, and therefore of $\rho$ itself,
needs at least eight states in its decomposition; this despite the
fact that the rank of $\rho$ is only six (see Lemma 1 of
Ref. \cite{barely}).  Thus we have a new example of a state for which
the number of states in its minimal decomposition exceeds its rank;
but see Ref. \cite{Lock}.  For general $d$, the number of states in
our separable ensemble at $B$, $3 {d\choose 2}$, which is more than the
dimension $d^2$ for $d > 3$.  There are no known prior explicit
examples in which the number of members of the optimal ensemble is
greater than the dimension; it would be interesting to prove that
Eq. (\ref{dec1}) constitutes a minimal optimal ensemble.

The separability of the PPT states permits us to give an extension of
Proposition \ref{abc1shotno} indicating that the undistillability of
states in this region is linked:

\begin{lem}
\label{momo2}
If the state $\rho_{bc}$ at point G is pseudo n-copy undistillable,
then all states in the region BGK are pseudo n-copy undistillable.
\end{lem}
{\em Proof:} First, note that if the state at point $G$ is pseudo
$n$-copy undistillable, then it is also pseudo $k$-copy undistillable
for $1\le k\le n$.  Since the two extremal points $B$ and $K$ of the
convex set of states $BGK$ are separable, the partial transpose of all
states in this region can be written as a convex combination (using
notation from Eq. (\ref{deco})):
\begin{equation}
\rho^{PT}=a_0\rho^{PT}_G+\sum_\alpha a_\alpha\Pi_\alpha,
\end{equation}
where $\Pi_\alpha$ are product projectors and $a_\alpha\ge 0$.
Applying Lemma \ref{x2n}, we consider the expectation value of $n$
copies of this state with respect to any Schmidt rank two vector
$\ket{v}$\cite{momo}:
\begin{equation}
\bra{v}(a_0\rho^{PT}_G+\sum_\alpha a_\alpha\Pi_\alpha)^{\otimes n}\ket{v}.
\label{bigme}
\end{equation}
We need to show that this is non-negative; we show this by
demonstrating that each term in the tensor product, when expanded out,
is not negative.  Consider a term containing $k$ $\rho^{PT}_G$ factors
and $n-k$ factors involving the projectors $\Pi_\alpha$.  We can apply
the $n-k$ projectors to $\ket{v}$; since they are all product
projectors, the projected vector $\ket{v'}$ still has Schmidt rank two
(or one).  So, the matrix element of Eq. (\ref{bigme}) is proportional
to
\begin{equation}
\bra{v'}(\rho^{PT}_G)^{\otimes k}\ket{v'}.
\end{equation}
But if $G$ is pseudo $k$ copy undistillable, this matrix element is
non-negative.$\Box$

Note that this analysis does not apply to state $C$, because the 
projectors $\Pi_{ij}$ of Eq. (\ref{deco}) are not product projectors;
therefore, they can increase the Schmidt rank of $\ket{v}$.

For $d=3$ we have performed extensive numerical studies to search for
states distillable by projection on two copies in the region $BCGK$.
We find none, reinforcing the indication of Lemma \ref{momo2} that
an entire region inside the ${\rm NPT}_1$ set will prove to be
undistillable.  The next section will provide further evidence for
this idea.

\section{Undistillability for multiple copies}
\label{nulls}

In this section we will obtain our strongest result, which suggests
that some of the NPT states $\rho_{bc}$ are not distillable.  We will
be able to conclude that for any finite $n$ there exists an $\eps$
such that $\rho(c,\eps)^{\otimes n}$ (Eq. (\ref{epsstates})) is not
entangled on any $2 \otimes 2$ subspace, and is therefore one-copy
undistillable.  This result can have only one of two further
implications: 1) For some $c$, this $\eps$ asymptotes to some finite
value $\bar{\eps}(c)$ as $n\rightarrow\infty$.  In this case, the NPT
states $\rho(c,\eps<\bar{\eps}(c))$ are absolutely undistillable.  2)
For all $c$, this $\eps$ goes to zero as $n\rightarrow\infty$.  In
this case all states immediately to the right of line $BK$ are
distillable; thus all $\rho_{bc}$ states with NPT would be
distillable, since all such states can be first mixed with some
separable $\rho_{bc}$ state (a LQ+CC operation) to bring it to the
$BK$ line.  But, one might say that the states near $BK$ are
``barely'' distillable: an arbitrarily large number of copies of the
state are required before there is any sign of undistillability of the
state.  It would be fair to say that these states would still be
undistillable in any practical sense.

First, we establish the significance of the null-space properties of
$\rho(c,\eps=0)$ for the argument.  We consider the function
\begin{equation}
f(c,\eps,n)=\min_{\ket{\psi^2}}\bra{\psi^2}\, (\rho^{PT}(c,\eps))^
{\otimes n}\,\ket{\psi^2}.\label{srank2neg}
\end{equation}
Here the minimum is taken over all Schmidt rank two states
$\ket{\psi^2}$ in the full $d^n\otimes d^n$ Hilbert space.  By Lemmas
\ref{horlem} and \ref{x2n}, we know that the sign of $f(c,\eps,n)$
determines whether $\rho(c,\eps)$ is pseudo n-copy undistillable.  For
$\eps=0$ the state is separable and therefore $f(c,\eps=0,n)\ge 0$ for
all $n$.  The question is, does the state become pseudo n-copy
undistilable as $\eps\rightarrow 0$?  The answer is provided by the
result whose proof we outline in a moment, that there is no Schmidt
rank two vector in the null space of $\rho^{PT}(c,\eps=0)^{\otimes n}$
for any $n$.  In other words, for all $n$ and $c$,
\begin{equation}
f(c,\eps=0,n)>0.
\end{equation}
And, since $f$ is a continuous function of $\eps$, there must therefore
exist an $\eps_0(c,n)>0$ such that
\begin{equation}
f(c,0\le\eps\le\eps_0(c,n),n)\ge 0.
\end{equation}
Thus, there is a finite range of $\eps>0$ for every $n$ (and for all
$c$ such that $0\le c<1/(d(d-1))$) such that $\rho(c,\eps)$ is
pseudo n-copy undistillable.

The only knowledge lacking at this point for a complete demonstration
of the undistillability of $\rho(c,\eps)$ is the asymptotic behavior
of $\eps_0(c,n)$ as $n\rightarrow\infty$.  If $\eps_0(n)\rightarrow 0$
as $n\rightarrow\infty$, then NPT undistillability would not be
established; we would merely have shown that distillation becomes
difficult as $\eps\rightarrow 0$, requiring more and more copies of
the state in the distillation protocol.  If $\eps_0(c,n)$ remains
larger than some positive $\bar{\eps}(c)$ for all $n$ and for some
$c$, then we would know that all states
$\rho(c,0\le\eps\le\bar{\eps}(c))$ are absolutely undistillable.
Since the signs from our few-copy work are that indeed this threshold
remains positive we are led to the conjecture:
\medskip

{\bf Conjecture:} {\em States $\rho(c,\eps)$ of Eq. (\ref{epsstates}),
for sufficiently small positive $\eps$, are undistillable.}
\medskip

We can further speculate that the undistillable region will correspond
exactly to region $BCGK$ in which the state is pseudo one-copy and,
apparently, pseudo two-copy undistillable.  It may well be that pseudo
one-copy undistillability and absolute undistillability are
equivalent.

Now we present our result about the null-space properties of
$\rho(c,\eps=0)$ on which the above discussion is based: its null
space does not contain any non-zero vectors of Schmidt rank less than
three.

First we set up some notation.  Plain roman indices take values from
$0$ to $d-1$ unless otherwise stated. Let indices with superscript $p$
represent composite indices, e.g. $i^{p}_1$ represents $(l_1,m_1
|l_1\ne m_1)$. Let indices with superscript $e$ represent plain
indices, e.g., $i^{e}_1= i_1$.  Label the eigenvectors
(Eq. (\ref{eigvecs1})) as $\ket{i_k^{e}}=\ket{\Phi_{i_k}}$ and
$\ket{i_k^{p}}=\ket{l_km_k}$ with $l_k\ne m_k$.  The label $e$ stands
for ``e-dit eigenstate'' and the label $p$ stands for ``product
eigenstate''. Let us denote $n$-tuples of indices such as
$(i_1,i_2,...,i_n)$ by letters in bold font such as {\bf i}; in sums
over {\bf i}, each $i_k$ runs independently between $0$ and $d-1$.

Next we prove an important lemma:
\begin{lem}
\label{LEM:FIRST}
The null space of 
the partial transpose of density matrix $\rho(c,\eps=0)^{\otimes n}$, for all
$c$, $d\ge3$ and  $n\ge 1$, does not contain any non-zero vectors with Schmidt
rank less than three of the form
\begin{equation}
\ket{\psi^{ee...e}} = \sum_{{\bf i}=0}^{d-1}
a_{\bf i} \ket{\Phi_{i_1}} \otimes \ket{\Phi_{i_2}}
\otimes ... \otimes \ket{\Phi_{i_n}} \enspace ,
\label{bigvec}
\end{equation}
\end{lem}
{\em Proof:} For $n=1$ the result is obvious since $\Phi_0$ is the
only vector in the null space and it has Schmidt rank $d\ge3$.  For
$n\ge 2$, we first note that the partial trace of the state in
Eq. (\ref{bigvec}) is
\begin{equation}
\rho_{\psi} ={\rm Tr}_B\proj{\psi^{ee...e}}=\sum_{{\bf i}=0}^{d-1}
|\tilde{a}_{{\bf i}}|^2
\proj{{\bf i}},
\end{equation}
where the coefficients
\begin{equation}
\tilde{a}_{{\bf i}}= \frac{1}{d^{(n/2)}}
\sum_{{\bf k}=0}^{d-1} a_{\bf k}\,
e^{i2\pi({\bf i} \cdot {\bf k})/d} \enspace
\end{equation}
are the $n$-dimensional discrete Fourier transforms of the $a$'s.
Here $ {\bf i} \cdot {\bf k}= i_1 k_1 + i_2 k_2 + ... + i_n k_n$.
Note that $\rho_{\psi}$ is already diagonalized. Since the
Schmidt rank of a pure state equals the rank of the partial trace,
we require that the rank of  $\rho_{\psi}$ be less than three.
Thus at most two coefficients $|\tilde{a}_{{\bf i}}|^2$ are nonzero, i.e.
\begin{equation}
\tilde{a}_{\bf i} =
|\alpha|\, e^{i\phi_\alpha} \delta_{{\bf i},{\bf x}} +
|\beta|\, e^{i\phi_\beta} \delta_{{\bf i},{\bf y}}
\enspace,
\end{equation}
where $\delta_{\bf i,x}= \delta_{i_1,x_1}
\delta_{i_2,x_2}...\delta_{i_n,x_n}$.

Solving for the $a_{\bf i}$s by doing an inverse Fourier transform we have
\begin{equation}
a_{\bf i} = \frac{1}{d^{(n/2)}}
(|\alpha| \,e^{i\phi_\alpha} \,e^{-i2\pi {\bf i}\cdot {\bf x}/d}  +
|\beta| \,e^{i \phi_\beta} \,e^{-i2\pi{\bf i} \cdot {\bf y}/d})\enspace.
\end{equation}

First suppose the Schmidt rank of the vector is exactly two, in which
case both $|\alpha|$ and $|\beta|$ must be nonzero and ${\bf x} \ne
{\bf y}$.  Now we start putting constraints on the $a$'s such that the
vector $\ket{\psi^{ee...e}}$ is in the null space of
$\rho^{PT}(c,\epsilon=0)$.  The vector $\psi^{ee...e}$ belongs to the
null space only if $a_{11...1}=0$, because the corresponding
eigenvalue $\lambda_1^n$ is positive at $\epsilon=0$.  Now we impose
the null space constraint $a_{21...1}=0$ (since the corresponding
eigenvalue $\lambda_2\lambda_1^{n-1}$ is positive at $\epsilon=0$),
and we have $x_1=y_1$. Similarly, other $a$'s, whose subscripts are
obtained by permuting $\{21...1\}$, may be constrained to zero giving
${\bf x}={\bf y}$. However this implies that the vector is of Schmidt
rank one if it is to satisfy these null space constraints.  Thus no
Schmidt rank two vector of the form $\psi^{ee..e}$ belongs to the null
space of $\rho^{PT}(c,\epsilon=0)$.

Next we consider the case of Schmidt rank one vectors, where without
loss of generality we may assume $|\beta|=0$. Then, the null space
constraint $a_{11...1}=0$ implies that $|\alpha|=0$, thus proving the
result.  $\Box$
\medskip

Now we are ready for the main result:
\begin{theo}
The null space of $(\rho^{PT}(c,\eps=0))^{\otimes n}$ for $d\ge 3$ and
$n\ge 1$ does not contain any vector of Schmidt rank less than three.
\end{theo}
{\em Proof:} For $n=1$ the result is obvious, because the null space
consists of the span of the vector $\ket{\Phi_0}$ which has Schmidt
rank $d\ge 3$.  For purpose of illustrating the proof technique, we
next prove the result for two copies, i.e., $n=2$.  Then we will show
how the proof generalizes to $n$ copies.

Recalling Eq. (\ref{eigvecs1}) and the fact that the eigenvectors
form a basis for the one-copy Hilbert space of $d \otimes d$, a general
vector $\ket{\psi}$ in the Hilbert space of two copies can be written as
\begin{equation}
\ket{\psi} = \ket{\psi^{ee}}+\ket{\psi^{ep}}+\ket{\psi^{pe}}+
\ket{\psi^{pp}} \enspace,
\label{twocop}
\end{equation}
with $\ket{\psi^{ee}}=\sum_{i^e,j^e} \alpha_{i^e,j^e}^{ee} \ket{i^e}
\otimes \ket{j^e}$, $\ket{\psi^{ep}} =\sum_{i^e,j^p} \alpha_{i^e
,j^p}^{ep} \ket{i^e} \otimes \ket{j^p}$, $\ket{\psi^{pe}}=\sum_{i^p,
j^e} \alpha_{i^p, j^e}^{pe} \ket{i^p} \otimes \ket{j^e}$ and
$\ket{\psi^{pp}}=\sum_{i^p,j^p} \alpha_{i^p,j^p}^{pp} \ket{i^p}
\otimes \ket{j^p}$. Here the $\alpha$'s are complex coefficients for
the vectors.  The $\psi^{pp}$ term must be zero if the vector is to
belong to the null space, because the corresponding eigenvalue
$\lambda_2^2$ is positive at $\epsilon=0$.  Now assuming $\psi$ has
Schmidt rank less than three, we will show that the coefficients
$\alpha^{ep}$'s and $\alpha^{pe}$'s are zero.  To show this we
repeatedly use the fact that local projections cannot increase the
Schmidt rank of a vector.  Alice and Bob can project locally on the
vector $\ket{k^p}$, for any $k^p$ of the first copy, which results in
a vector proportional to $\ket{k^p} \otimes \sum_{j^e} \alpha_{k^p
j^e}^{pe} \ket{j^e}$.  By Lemma \ref{LEM:FIRST} this vector has
Schmidt rank greater than two unless it is zero. Thus all the
$\alpha^{pe}$'s are zero. Similarly applying this argument to the
$\psi^{ep}$ term, with the projection now done on a product vector
$\ket{k^p}$ of the second copy, we see that the $\alpha^{ep}$'s are
zero. The only term left now is the $\psi^{ee}$ term, for which Lemma
\ref{LEM:FIRST} applies and gives us the result.

We write the general proof for $n$ copies along the lines of the
two-copy proof, albeit with considerable notational complications.
Generalizing the notation of Eq. (\ref{twocop}), we define ${\cal
P}_k$ to be the set of all distinct permutations of $k$ $p$'s and
$(n-k)$ $e$'s.  We also denote the strings representing permutations
in ${\cal P}_k$ by bold font, e.g., ${\bf s}=s_1s_2...s_k$, where the
$s_j$ are the characters in the permutation string, e.g., for ${\bf
s}=pep\in{\cal P}_2$, then $s_1=p$, $s_2=e$, and $s_3=p$.

A general state in the $n$-copy Hilbert space can be written in the form
\begin{equation}
\ket{\psi} = \sum_{k=0}^{n}\sum_{s\in {\cal P}_k}\ket{\psi^{\bf s}} \enspace,
\end{equation}
with
\begin{equation}
\ket{\psi^{\bf s}}= \ket{\psi^{s_1s_2...s_n}}=
\sum_{i_1^{s_1},i_2^{s_2},...,i_n^{s_n}}
\alpha_{i_1^{s_1},i_2^{s_2},...,i_n^{s_n}}
\ket{i_1^{s_1}} \otimes \ket{i_2^{s_2}} \otimes ... \otimes \ket{i_n^{s_n}}
\enspace.
\end{equation}
Again, the $\psi^{pp...p}$ term is zero if the vector is to be in the
null space, because the corresponding eigenvalue $\lambda_2^n$ is positive
at $\epsilon=0$.
Define $\psi_m$ by
\begin{equation}
\ket{\psi_m} = \sum_{i=0}^m\sum_{{\bf s}\in {\cal P}_i}\ket{\psi^{\bf s}}
\enspace.
\end{equation}
Then to prove the result we show that there is no vector with Schmidt
rank less than three
of the form $\ket{\psi_l}$ for all $m\le n-1$. This we show by
induction on $m$.  For $m=0$, the result immediately follows
from Lemma \ref{LEM:FIRST}. For the induction step, we write
\begin{equation}
\ket{\psi_m} = \ket{\psi_{m-1}} + \sum_{{\bf s}
\in {\cal P}_m} \ket{\psi^{\bf s}}
\enspace.
\end{equation}
Now if Alice and Bob locally project $\ket{\psi_m}$ onto $\ket{r_i^p}$
of the $i$th copy, for $i=1...m$, the result is
\begin{equation}
\ket{r_1^p}\otimes \ket{r_2^p} \otimes ...\otimes \ket{r_m^p}
\otimes \sum_{k^e_{m+1} ... k^e_n}
\alpha_{r_1^p r_2^p ...r_m^p k_{m+1}^e ... k_n^e}
\ket{k_{m+1}^e} \otimes ... \otimes \ket{k_n^e} \enspace.
\end{equation}
Since local projection cannot increase the Schmidt rank, by Lemma
\ref{LEM:FIRST} the vector inside the sum above must be zero.  Doing
this for all the different values of the $r_i^p$'s we see that
$\psi^{ppp...pee...e}=0$, where the superscript contains $m$ $p$'s and
$(n-m)$ $e$'s.  Similarly we can prove that $\ket{\psi^{\bf s}}$ is
zero for any permutation string ${\bf s} \in {\cal P}_m$.  This shows
that $\ket{\psi_m}$ has to be of the form $\ket{\psi_{m-1}}$, for
which the result is true by the induction hypothesis.  $\Box$
\medskip

\section{Distillability and $2$-positive linear maps}
\label{twopos}

In this section we find a formulation of the problem of distillability
of an arbitrary bipartite density matrix $\rho$. This formulation uses
the notion of $2$-positive linear maps. We will explicitly show how
the problem of distillability of the density matrices $\rho_{bc}$ that
were discussed in the preceding sections can be cast in the language
of positive linear maps.

Let us first recall the definition of a $k$-positive linear map
\cite{posmaplit}. Let $B({\cal H}_n)$ denote the matrix algebra of
operators on an $n$-dimensional Hilbert space and let $B({\cal
H}_n)^+$ denote the set of positive semidefinite matrices. A
linear map $\Lambda\, \colon B({\cal H}_n) \rightarrow B({\cal H}_m)$
is called positive when $\Lambda \,\colon B({\cal H}_n)^+ \rightarrow
B({\cal H}_m)^+$, that is, the map preserves the set of positive
semidefinite matrices.  A linear map $\Lambda\, \colon B({\cal H}_n)
\rightarrow B({\cal H}_m)$ is called $k$-positive when the map ${\bf
1}_k \otimes \Lambda \,\colon B({\cal H}_k \otimes {\cal H}_n)
\rightarrow B({\cal H}_k \otimes {\cal H}_m)$ is positive. Note that
$1$-positivity is equivalent to positivity. It is not hard to show
that when a map ${\Lambda}\, \colon B({\cal H}_n) \rightarrow B({\cal
H}_m)$ is $n$-positive, it is completely positive.

We will now give an alternative characterization of $k$-positivity.
The next lemma says that to test a linear map for $k$-positivity we only need
to apply it to pure states of at most Schmidt rank $k$.

\begin{lem}
A positive linear map $\Lambda \,\colon B({\cal H}_n) \rightarrow
B({\cal H}_m)$ is $k$-positive if and only if
\begin{equation}
({\bf 1}_n \otimes \Lambda)(\ket{\psi} \bra{\psi}) \geq 0,
\label{kposeq}
\end{equation}
for all vectors $\ket{\psi} \in {\cal H}_n \otimes {\cal H}_n$ which
have Schmidt rank at most $k$.
\label{kpos}
\end{lem}

{\em Proof:} If Eq. (\ref{kposeq}) holds for all states $\ket{\psi}$
of Schmidt rank at most $k$, then it follows that $({\bf 1}_k
\otimes\Lambda)(\ket{\psi} \bra{\psi}) \geq 0$ for all vectors
$\ket{\psi}\in {\cal H}_k \otimes {\cal H}_n$. Therefore $({\bf 1}_k
\otimes \Lambda)(\rho) \geq 0$ for all $\rho \in B({\cal H}_k \otimes
{\cal H}_n)^+$ and thus $\Lambda$ is $k$-positive. On the other hand,
if there exists a vector $\ket{\psi}$ of at most Schmidt rank $k$ for
which $({\bf 1}_n \otimes \Lambda)(\ket{\psi}\bra{\psi}) \ngeq 0$,
then $\Lambda$ cannot be $k$-positive. $\Box$

We would like to make an additional simplification in characterizing
$2$-positive maps. The next lemma says that in order to test a linear
map for $2$-positivity we only need to apply it to maximally entangled
pure states of Schmidt rank two.

\begin{lem}
A linear positive map $\Lambda \,\colon B({\cal H}_n) \rightarrow
B({\cal H}_m)$ is $2$-positive if and only if, for all
$\ket{\Psi^\beta}=\ket{0,\beta_0}+\ket{1,\beta_1}$ with
$\langle\beta_0|\beta_1\rangle=0$, $\langle\beta_0|\beta_0\rangle=
\langle\beta_1|\beta_1\rangle=1$,
\begin{equation}
({\bf 1}_2 \otimes \Lambda)(\ket{\Psi^\beta} \bra{\Psi^\beta})\geq 0.
\label{2posmax}
\end{equation}
\label{PROP2POS}
\end{lem}

The proof of this lemma is given in Appendix \ref{prooflem}.
It is possible to formulate a similar lemma for $k$-positive maps,
in which $k$-positivity or the lack thereof can be deduced from
applying the map on all maximally entangled vectors of Schmidt rank $k$.
\vspace{0.5cm}

With a Hermitian operator $H \in B({\cal H}_d\otimes {\cal H}_d)$ we
can always associate a hermiticity-preserving linear map $\Lambda$ in
the following way:
\begin{equation}
H=({\bf 1}_d \otimes \Lambda)(\ket{\Phi^+}\bra{\Phi^+}).
\end{equation}
where
\begin{equation}
\ket{\Phi^+}=\frac{1}{\sqrt{d}} \sum_{i=0}^{d-1} \ket{ii}.
\end{equation}
In the Appendix of \cite{filterhor} it was proved that the operator
$H$ is positive semidefinite if and only the linear map $\Lambda$ is
completely positive.  From this we conclude that any bipartite density
matrix $\rho$ on $d \otimes d$ can always be written as
\begin{equation}
\rho=({\bf 1}_d \otimes {\cal S})(\ket{\Phi^+}\bra{\Phi^+}),
\label{rhocp}
\end{equation}
where ${\cal S}\,\colon B({\cal H}_d) \rightarrow B({\cal H}_d)$ is a
completely positive map.  Note that ${\cal S}$ need not be trace
preserving.

As an example, we derive the completely positive map ${\cal S}_{bc}$
associated with the density matrices $\rho_{bc}$ given in
Eq. (\ref{abcform}).  We can specify ${\cal S}_{bc}$ on the input
states:
\begin{equation}
\ba{lr}
{\cal S}_{bc}(\ket{i}\bra{i})=a\ket{i} \bra{i}+ \frac{b+c}{2}\sum_
{j \neq i} \ket{j}
\bra{j}, &
{\cal S}_{bc}(\ket{i}\bra{j})=\frac{c-b}{2}\ket{j}\bra{i},\;\; i \neq j.
\ea
\label{sexamp}
\end{equation}

The following main theorem expresses the connection between
$2$-positivity and distillability of a density matrix $\rho$:

\begin{theo}
Let $\rho$ be a bipartite density matrix on $d \otimes d$. Let ${\cal
S}\,\colon B({\cal H}_d) \rightarrow B({\cal H}_d)$ be a completely
positive map which is uniquely determined by
\begin{equation}
\rho=({\bf 1}_d \otimes {\cal S})(\ket{\Phi^+}\bra{\Phi^+}).
\label{rhoins}
\end{equation}
Let ${\Lambda}\, \colon B({\cal H}_d) \rightarrow B({\cal H}_d)$ be a
linear positive map defined as
\begin{equation}
\Lambda=T \circ {\cal S},
\label{defL}
\end{equation}
where $T$ is matrix transposition in the basis
$\{\ket{i}\}_{i=0}^{d-1}$.  There exists no projections $P_A\,\colon
{\cal H}_d^A \rightarrow {\cal H}_2$ and $P_B\,\colon {\cal H}_d^B
\rightarrow {\cal H}_2$ such that $(P_A \otimes P_B)\, \rho\, (P_A^{\dagger}
\otimes P_B^{\dagger})$ is entangled if and only if the map $\Lambda$ is
2-positive.  Let
\begin{equation}
\Lambda^{\otimes n}=\underbrace{\Lambda \otimes \ldots \otimes \Lambda}_{n}.
\label{compos}
\end{equation}
The density matrix $\rho$ is not distillable if and only if for all
$n=1,2,\ldots$ the map $\Lambda^{\otimes n}$ is 2-positive.
\label{2posing}
\end{theo}

{\em Proof:}
We will prove the theorem in two parts. First we will prove the relation
between $2$-positivity of $\Lambda$ and the nonexistence of a $2 \otimes 2$
subspace on which $\rho$ is entangled. Then we prove the result relating
undistillability to $2$-positivity of $\Lambda^{\otimes n}$.

Let us assume that there does not exist a $2 \otimes 2$ subspace on
which the density matrix $\rho$ is entangled.  We can write any
projector $P_A:\,{\cal H}_d \rightarrow {\cal H}_2$ as
\begin{equation}
P_A=\ket{0}\bra{\alpha_0}+\ket{1}\bra{\alpha_1},
\end{equation}
where $\bra{\alpha_0} \alpha_1 \rangle=0$. Lemma \ref{x2n} implies that
\begin{equation}
({\bf 1}_2 \otimes T)\left[(P_A \otimes {\bf 1}_d)\, \rho\, (P_A^\dagger
\otimes {\bf 1}_d)\right] \geq 0,
\label{ptpa}
\end{equation}
for all projectors $P_A$. This expression, using the
Eqs. (\ref{rhoins}) and (\ref{defL}), is equal to
\begin{equation}
({\bf 1}_2 \otimes \Lambda)(\ket{\Psi^{\alpha^*}} \bra{\Psi^{\alpha^*}})
\geq 0,
\label{shouldbepos}
\end{equation}
with $\ket{\Psi^{\alpha^*}} \in {\cal H}_2 \otimes {\cal H}_d$ defined as
\begin{equation}
\ket{\Psi^{\alpha^*}}=\frac{1}{\sqrt{2}}\left(\ket{0,\alpha_0^*}+
\ket{1,\alpha_1^*}\right),
\end{equation}
The vectors $\ket{\alpha_{0,1}^*}$ are defined as
$\ket{\alpha_{0,1}^*}=\sum_{i=0}^{d-1} \langle \alpha_{0,1} \ket{i}
\ket{i}$.  Note that $\langle \alpha_0^*\,|\, \alpha_1^*\rangle=0$.
We now invoke the property of a 2-positive map as given in Lemma
\ref{PROP2POS}; if Eq. (\ref{shouldbepos}) holds for all
$\ket{\alpha_0^*},\ket{\alpha_1^*}\in {\cal H}_n$ with $\langle
\alpha_0^*| \alpha_1^*\rangle=0$, then $\Lambda$ is a 2-positive
map. Conversely, invoking Lemma \ref{PROP2POS}, if $\Lambda$ is a
2-positive linear map, then Eq. (\ref{shouldbepos}) holds for all
states $\ket{\Psi^{\alpha^*}}$.  This implies that Eq. (\ref{ptpa})
holds for all projectors $P_A$ and thus there does not exist a $2
\otimes 2$ subspace on which $\rho$ is entangled.

Now we turn to the second part of the proof. The necessary and
sufficient condition for distillability of a density matrix was given
in Lemma \ref{horlem}.  Let $\rho^{\otimes n}=\rho \otimes \rho
\otimes \ldots \otimes \rho$ on $d^n \otimes d^n$.  The density matrix
$\rho$ is undistillable if and only if there exists no projections
$P_A\,\colon {\cal H}_{d^n}^A \rightarrow {\cal H}_2$ and $P_B\,\colon
{\cal H}_{d^n}^B \rightarrow {\cal H}_2$ such that $(P_A \otimes
P_B)\, \rho^{\otimes n}\, (P^\dagger_A \otimes P^\dagger_B)$ is
entangled.  Thus if a density matrix is undistillable, we have,
similar as Eq. (\ref{ptpa}),
\begin{equation}
({\bf 1}_2 \otimes T)\left[(P_A \otimes {\bf 1}_{d^n})\, \rho^{\otimes n}\,
(P^\dagger_A \otimes {\bf 1}_{d^n})\right] \geq 0,
\label{ptpak}
\end{equation}
for all projectors $P_A\,\colon {\cal H}_{d^n}^A \rightarrow {\cal
H}_2$ and all $n=1,2,\ldots$. We use the fact that $T\,\colon B({\cal
H}_d^{\otimes n}) \rightarrow B({\cal H}_d^{\otimes n})$ is equivalent
(up to a unitary transformation) to $T_d^{\otimes n}$ where $T_d$ is
matrix transposition in ${\cal H}_d$.  Then Eq. (\ref{ptpak}) can be
rewritten as
\begin{equation}
({\bf 1}_2 \otimes \Lambda^{\otimes n})(\ket{\Psi} \bra{\Psi}) \geq 0,
\end{equation}
for all maximally entangled states $\ket{\Psi} \in {\cal H}_2 \otimes
{\cal H}_{d^n}$ for all $n=1,2,\ldots$.  This implies with Lemma
\ref{PROP2POS} that $\Lambda^{\otimes n}$ is $2$-positive for all
$n=1,2,\ldots$.  Conversely, when $\Lambda^{\otimes n}$ is not
$2$-positive for some $n$, there will exist a $2 \otimes 2$ subspace
on which $\rho^{\otimes n}$ is entangled. $\Box$

{\em Remarks:} Note that the theorem also holds for entangled density matrices
$\rho$ that have the PPT property or density matrices which are separable. In this
case, however, the positive map $\Lambda$ is completely positive, and
therefore the map $\Lambda^{\otimes n}$ for all
$n=1,2,\ldots$ is $2$-positive trivially.

We note that Theorem \ref{2posing} can also be made to apply to a
situation in which one is given a large number of copies of, say, two
different density matrices $\rho_1$ and $\rho_2$. With each of these
density matrices we associate a positive linear map $\Lambda_1$ and
$\Lambda_2$.  Distillability of $\rho_1$ and $\rho_2$ together can be
formulated as the problem of determining whether $\Lambda_1^{\otimes
n_1} \otimes \Lambda_2^{\otimes n_2}$ is $2$-positive.  This provides
a method for searching for nonadditivity in the property of
distillability\cite{active}.  We could encounter a situation in which
both $\rho_1$ and $\rho_2$ are undistillable, but $\rho_1$ and
$\rho_2$ taken together are distillable.

In general, given two $2$-positive maps $\Lambda_1$ and $\Lambda_2$,
the tensor product $\Lambda_1 \otimes \Lambda_2$ is not necessarily
$2$-positive.  As an example we take $\Lambda_1$ to be the identity
map ${\bf 1}_d$ and $\Lambda_2$ a $2$-positive map which is not
$2d$-positive. Then by definition, ${\bf 1}_2 \otimes {\bf 1}_d
\otimes \Lambda_2$ is not positive. In the cases that we consider here
however, the maps are of a special form, namely $\Lambda=T \circ {\cal S}$,
where ${\cal S}$ is completely positive.  For this special form,
it is possible that the composed maps are always 2-positive.

The positive map $\Lambda_{bc}$ of Eq. (\ref{compos}) corresponding to
the example Eq. (\ref{sexamp}) is
\begin{equation}
\ba{lr}
\Lambda_{bc}(\ket{i}\bra{i})=a\ket{i} \bra{i}+ \frac{b+c}{2}\sum_{j \neq i}
\ket{j}\bra{j}, &
\Lambda_{bc}(\ket{i}\bra{j})=\frac{c-b}{2}\ket{i}\bra{j},\;\;i \neq j.
\ea
\end{equation}
For the states on the line $FH$ this corresponds to the positive map
$\tau_W$ which acts as
\begin{equation}
\tau_{W}(X)=d\lambda{\bf 1}{\rm Tr} X -(\lambda+1)X,
\end{equation}
where $\lambda$ is the parameter in Eq. (\ref{ldef}).

It has been shown \cite{uhlpriv} that this map $\tau_W$ is 2-positive
in the region $\lambda \geq \frac{2}{d-2}$. This thus establishes an
alternative proof of Theorem \ref{1shot} in section \ref{single}.

\section{Conclusion}

Our alternative formulation of the problem of distillability in terms
of the $2$-positivity property of linear maps has not yet led to a
solution of the problem of NPT density matrices which are (likely to
be) undistillable (Conjecture at the end of Sec. \ref{nulls}).  We
present the formulation here, as it points to a new connection between
the structure of positive linear maps and the classification of
bipartite mixed state entanglement.  We expect that fruitful results
will flow from understanding in more detail the classification schemes
for these NPT states that are based directly on their 2-positivity
properties.

In conclusion, we have shown that most of the
distillability properties of NPT mixed states can be restricted to the
study of the canonical set $\rho_{bc}$.  Many of the questions about
one-copy and few-copy distillability of these states are completely
answered by our analysis.  A final, general proof of the full
undistillability of these states eludes us, but we have shown that if
they are distillable, it involves a much more difficult protocol than
any which has been needed up until now.

{\em Note added:} After the completion of the calculations reported
here, we became aware of closely related work by D{\"u}r {\em et
al.}\cite{Du}.  This paper studies the states along the line {\em HGF}
in Fig. \ref{NPTreg}; for these states it provides an alternative
proof to the one discussed here in Sec. \ref{nulls} that, approaching
point $H$, the states are pseudo $n$-copy undistillable for any $n$.
Ref. \cite{Du} also obtains the same theorem as here (our Theorem
\ref{1shot}) about pseudo one-copy distillability of these states, as
well as obtaining additional numerical results indicating that the
region of pseudo two- and three-copy undistillability is the same as
that for one-copy undistillability.  All the results of Ref. \cite{Du}
and the present work are consistent.

\section*{Acknowledgments}

DPD, JAS, and BMT acknowledge support from the Army Research Office
under contract number DAAG55-98-C-0041.  AVT acknowledges support from
the Army Research Office under contract number DAAG55-98-1-0366.  PWS
is supported in part by DARPA through Caltech's Quantum Information
and Computation (QUIC) project administered by the Army Research
Office under Grant No. DAAH04-96-1-0386.  We are grateful to R. Jozsa
and M. Jeng for permission to use several of their results in this
paper.  We thank Charles Bennett, Micha\l{} Horodecki, Pawe\l{}
Horodecki, and Armin Uhlmann for interesting discussions.

\appendix

\section{Proof of Lemma \protect\ref{PROP2POS}}
\label{prooflem}

The proof is similar in structure to the proof of the lemma in the
Appendix of \cite{filterhor}.  By definition a linear positive map
$\Lambda\,\colon B({\cal H}_n) \rightarrow B({\cal H}_m)$ is
2-positive if and only if, for {\em all} $\ket{\psi} \in {\cal H}_2
\otimes {\cal H}_n$,
\begin{equation}
({\bf 1}_2 \otimes \Lambda)(\ket{\psi} \bra{\psi}) \geq 0.
\label{anystate}
\end{equation}
We will show that we only need to consider states $\ket{\psi}$ that
are maximally entangled.  Note that any (unnormalized) maximally
entangled state can be written as
$\ket{\Psi^\beta}=\ket{0,\beta_0}+\ket{1,\beta_1}$ with
$\langle\beta_0| \beta_1\rangle=0$,
$\langle\beta_0|\beta_0\rangle=\langle\beta_1|\beta_1\rangle=1$.
We start with the following observation: When we apply the map
$\Lambda$ on some maximally entangled state in ${\cal H}_2 \otimes
{\cal H}_n$,
\begin{equation}
D=({\bf 1}_2 \otimes \Lambda)(\ket{\Psi^\beta} \bra{\Psi^\beta}),
\label{defD}
\end{equation}
the matrix $D$ uniquely determines the action of the map $\Lambda$ on
any input matrix that has support on the two dimensional space spanned
by the vectors $\ket{\beta_0}$ and $\ket{\beta_1}$.

For the first part of the Lemma, let $D \geq 0$ in Eq. (\ref{defD}). Since $D$ is Hermitian, we can
write it in its eigendecomposition
\begin{equation}
D=\sum_i \mu_i \ket{\phi_i} \bra{\phi_i},
\end{equation}
with the eigenvalues $\mu_i \geq 0$ and the eigenvectors
$\ket{\phi_i}\in {\cal H}_2 \otimes {\cal H}_n$.  Each eigenstate
$\ket{\phi_i}$ can be written in a Schmidt decomposition as
$\ket{\phi_i}=\sqrt{\lambda_{0,i}}\ket{\alpha_{0,i},\beta_{0,i}}+
\sqrt{\lambda_{1,i}}\ket{\alpha_{1,i},\beta_{1,i}}$
with $\langle \beta_{0,i} \ket{\beta_{1,i}}=\langle \alpha_{0,i}
\ket{\alpha_{1,i}}=0$, and all vectors normalized.  Note that the states
$\ket{\beta_{0,i}}$ and $\ket{\beta_{1,i}}$ can span a different
two-dimensional subspace of ${\cal H}_n$ for each $i$.  There exists a
local filter $W_i^\beta$
\cite{filterhor}
from which we can
obtain the state $\ket{\phi_i}$ from the maximally entangled state
$\ket{\Psi^\beta}$:
\begin{equation}
\ket{\phi_i} \bra{\phi_i}=({\bf 1}_2 \otimes W_i^\beta)
(\ket{\Psi^\beta}\bra{\Psi^\beta})({\bf 1}_2 \otimes {W_i^\beta}
^{\dagger}),
\end{equation}
$W_i^\beta$ includes: (1) a unitary transformation from the basis
$\beta^\prime_{(0,1),i}$ to $\beta_{(0,1),i}$, where $\beta^\prime$
are the Schmidt vectors of $\Psi^\beta$ when it is written in
the form $\ket{\Psi^\beta}=\ket{\alpha_{0,i},\beta^\prime_{0,i}}
+\ket{\alpha_{1,i},\beta^\prime_{1,i}}$ (taking advantage of the
degeneracy of the Schmidt decomposition of the maximally
entangled state), and (2) a diagonal filter which reduces the
Schmidt coefficients to $\lambda_{(0,1),i}$.

Thus we may write $D$ as
\begin{equation}
D=\sum_i \mu_i\, ({\bf 1}_2 \otimes W_i^\beta)\, \ket{\Psi^\beta}
\bra{\Psi^\beta}\, ({\bf 1}_2 \otimes {W_i^\beta}^\dagger).
\end{equation}
We see that since $D \geq 0$ by assumption, we are
able to write the action of the map $\Lambda$ on the input
$\ket{\Psi^\beta}$ in a `completely positive form' with operation
elements $\sqrt{\mu_i}W_i^\beta$ that depend on $\beta$.  We observed
above that this input determines the action of the map uniquely on the
subspace spanned by the vectors $\ket{\beta_0}$ and $\ket{\beta_1}$.
Therefore the map acts as a completely positive map on any input that
has support on a two-dimensional space. This implies that
Eq. (\ref{anystate}) holds for any state $\ket{\psi} \in {\cal
H}_2\otimes {\cal H}_n$.  Conversely, if $\Lambda$ is $2$-positive
then Eq. (\ref{2posmax}) holds for any maximally entangled state
$\ket{\Psi^\beta}$. $\Box$
\bigskip
\bigskip

\begin{references}
\bibitem{foot1} The notation $n\otimes m$ means that Alice's Hilbert
space ${\cal H}_n$ has dimension $n$, and Bob's Hilbert space ${\cal
H}_m$ has dimension $m$.
\end{references}
\end{document}